\documentclass{ws-style}

\begin{document}

\markboth{Robert Lauer for the IceCube Collaboration}
{Extended search for neutrino point sources with IceCube}

\title{Extended search for point sources of neutrinos below and above the horizon: Covering energies from TeV to EeV with IceCube}

\author{Robert Lauer for the IceCube collaboration\footnote{See http://www.icecube.wisc.edu for a full list of authors}}
\address{DESY, Platanenallee 6,\\
Zeuthen, 15738, Germany\\
robert.lauer@desy.de}
\maketitle
\begin{abstract}
Point source searches with neutrino telescopes like IceCube are normally restricted to one hemisphere, due to the selection of up-going events as a way of rejecting the atmospheric muon background.
In this work we show that the down-going region above the horizon can be included in the search by suppressing the background through energy-sensitive selection procedures. This approach increases the reach to the EeV regime of the signal spectrum, which was previously not accessible due to the absorption of neutrinos with energies above a PeV inside the Earth.
We present preliminary results of this analysis, which for the first time includes up-going as well as down-going muon events in a combined approach. We used data collected with IceCube 
in a configuration of 22 strings. No significant excess above the atmospheric background is observed. 
While other analyses provided results for the Northern hemisphere, this new approach extends the field of view to a large part of the Southern sky, which was previously not covered with IceCube.
\end{abstract}
\section{Introduction}
In a neutrino point source analysis one focuses on obtaining the best angular resolution by observing muon tracks inside the detector volume. The expected signal stems from charged-current interactions of muon neutrinos and a small fraction of tau neutrinos.
The background consists primarily of high energy muons, produced in the atmosphere in extended air showers, with a much lower but irreducible contribution of atmospheric neutrinos from the same interactions. In IceCube~[\refcite{icecube}], Cherenkov light from these events is detected in digital optical modules, with 60 of these arranged vertically on a string. A total of 80 strings will be installed until 2011, situated at a depth of 1.45 to 2.45~km inside the glacial ice at the geographic South Pole.
A standard method to suppress atmospheric muons, reaching the detector from above, is a cut on the reconstructed track direction at the horizon, which allows IceCube to cover only the Northern hemisphere, as presented in [\refcite{jlbazoPS08}]. Inside the Earth, the neutrino-nucleus cross section rises with energy and prevents the majority of up-going neutrinos with energies above a PeV to reach the vicinity of the detector.
\pagebreak
In our extended analysis, down-going tracks are included and their energy is used to separate signal neutrinos from atmospheric background, based on the softer spectrum of the latter. Different zenith dependent cuts make it possible to perform a point source search above and below the horizon.

\section{Event selection and reconstruction}
The analysis presented here used IceCube data collected between May 2007 and April 2008 (276 days) with 22 strings. An on-line filter stream provided an all-sky muon track sample with a rate of $23.6$ Hz, and elaborate direction reconstructions were performed off-line. Down-going events with declinations smaller than $-50^{\circ}$ were rejected due the small interaction volume above the detector leading to very low signal expectations. 
The steeply rising muon background above the horizon is dominated by multiple muons forming so-called bundles. To separate them from neutrino-induced single muon tracks, a selection cut based on the resolution of individual photon pulses in an optical module was applied. The threshold values were a function of the reconstructed zenith angle and energy for each event.
Employing parameters depending on the quality of the fitted track, similar to those in [\refcite{jlbazoPS08}], the background was reduced further. By systematically testing combinations of threshold values in dependence of zenith angle, these cuts were optimized for the best sensitivity considering two template spectra, a standard $E^{-2}$ and a harder $E^{-1.5}$ power law of the neutrino energy $E$.\\
The resulting sample is composed of 1877 events between $-50^{\circ}$ and $+85^{\circ}$. Simulations show that the down-going events from the Southern hemisphere are dominated by atmospheric muons, while the up-going tracks are mainly induced by atmospheric neutrinos.
Related to zenith dependent signal and background expectations, the cuts allow for an excess of events in the region just above the horizon.
The neutrino signal efficiency, as estimated from simulations, improves with energy and effectively defines a threshold rising from the TeV regime in the Northern hemisphere to PeV energies in the South. 
Compared to previous point source analyses in IceCube~[\refcite{jlbazoPS08}], an improved direction reconstruction was used. 
It is a likelihood algorithm that uses not only the timing information of the first pulse in each optical module, but also the total charge to account for the probability of multiple photoelectrons from individually scattered photons.
This is particularly important for neutrino energies above 100 TeV, where photons from secondaries produced along the muon track are an important addition to the light yield. 
The median angular resolution for the final event selection is $1.3^{\circ}$ ($1.2^{\circ}$) under the assumption of an $E^{-2}$ ($E^{-1.5}$) spectrum.

\begin{figure}[pt]
\centerline{\psfig{file=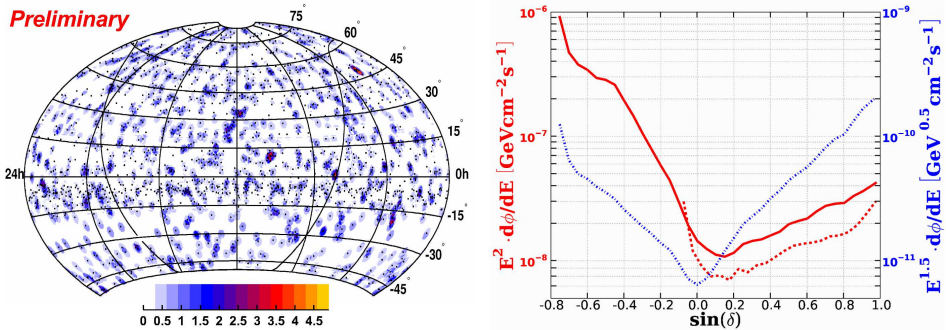,height=4.5cm}}
\vspace*{8pt}
\caption{Left: Sky map with bin significances, i.e. pre-trial {-$\log$(p-value)}. Right: Sensitivities from this work versus sinus of declination for two spectra, $E^{-2}$ (solid red) and $E^{-1.5}$ (dash-dotted blue), and for $E^{-2}$ the comparison to the up-going analysis (dashed red) from [2]. \label{f1}}
\end{figure}

\section{Point source search method and results}
We performed a binned scan of the entire sky map and also tested a list of pre-defined directions, with a search bin radius of $2.5^{\circ}$.
Details on the method for the all-sky search are given in [\refcite{ackermann5Y07}]. The major difference is that in this work we extended the scan to the Southern sky.
No significant deviation from the background-only hypothesis was observed. The fluctuation with the highest significance was found in the direction of dec. $1.0^{\circ}$ and r.a. $103.5^{\circ}$ (8 events observed, 1.2 expected). By performing searches on randomized sky maps, we determined the probability for an equally or more significant excess anywhere in the sky due to a background fluctuation to be 37\%.\\
In the second test a pre-defined list of source candidates was used, aiming at reducing the effects of trial factors.
From the confirmed Active Galactic Nuclei (AGN) listed in the third EGRET catalog~[\refcite{3EG99}] we selected 23 bright blazars by imposing thresholds for average (maximum) photon fluxes of $15\!\times\!10^{-8}$ ($40\!\times\!10^{-8}$) ph cm$^{-2}$s$^{-1}$. Two additional blazars reported 
in November 2008 were also included, see [\refcite{atel1}],[\refcite{atel2}], as well as two close-by AGN, M87 and Centaurus A,
and the center of our galaxy, Sgr A*.
No significant excess above background was found for any of these candidates. The upward fluctuation with the highest significance occurs for PKS 1622-297 with 1 event observed and 0.3 expected. From background simulation using randomized data, we expect a similar or more significant result from any of the sources in 98\% of all cases. The limits given in table~\ref{sourcelist}, calculated according to~[\refcite{fc}], are preliminary and do not yet include systematic uncertainties.
A previous point source search with better sensitivity for the Northern sky has been performed, and it was decided a priori that it gives the final results for the up-going region (dec. $>-5^{\circ}$) [\refcite{jlbazoPS08}].

\section{Summary and outlook} 
A new approach to neutrino point source searches extends both below and above the horizon and increases the reach of IceCube above PeV energies and to the Southern sky. Results from data taken with 22 strings are consistent with the background-only hypothesis. Details will be discussed in an upcoming publication. IceCube is taking data with 40 strings since April 2008 and has been recently extended to 58 strings, with better sensitivity to be expected for upcoming analyses.
\footnotetext[1]{For this source, flux limits from the up-going search will be shown in an upcoming publication.}

\begin{table}[ph]
\tbl{Results for pre-defined source candidates. N$_{90}$ is the neutrino event upper limit of the Feldman-Cousins 90\% confidence interval, $\Phi_{90}^{-\gamma}$ the resulting flux upper limit for an $E^{-\gamma}$ neutrino spectrum, i.e. $d\Phi/dE\leq \Phi_{90}^{-\gamma} 10^{-9} \text{GeV}^{-1}\text{cm}^{-2}\text{s}^{-1}(E/\text{GeV})^{-\gamma}$ with $\gamma=2$ or $1.5$. $\Delta E^{-\gamma}$ are the energy ranges of 90\% signal containment.}
{\begin{tabular}{@{}lccccccc@{}} \toprule
Object & r.a.[$^{\circ}$] & dec.[$^{\circ}$] & N$_{90}$ & $\Phi_{90}^{-2}$ & $\Delta E^{-2}$ [GeV] & $10^3\Phi_{90}^{-1.5}$& $\Delta E^{-1.5}$ [GeV]  \\ \colrule
 PKS 0537-441 & 84.7 & -44.1 & 1.7 & 261.3 & $3\!\!\times\!\!10^{6}$ - $7\!\!\times\!\!10^{8}$ & 35.4 & $7\!\!\times\!\!10^{6}$ - $6\!\!\times\!\!10^{9}$ \\ 
 Centaurus A & 201.4 & -43.0 & 3.7 & 541.4 & $2\!\!\times\!\!10^{6}$ - $8\!\!\times\!\!10^{8}$ & 72.6 & $7\!\!\times\!\!10^{6}$ - $6\!\!\times\!\!10^{9}$ \\
PKS 1454-354 & 224.4 & -35.6 & 1.9 & 218.5 & $1\!\!\times\!\!10^{6}$ - $9\!\!\times\!\!10^{8}$ & 29.1 & $6\!\!\times\!\!10^{6}$ - $7\!\!\times\!\!10^{9}$ \\
 PKS 1622-297 & 246.5 & -29.9 & 4.1 & 439.2 & $1\!\!\times\!\!10^{6}$ - $9\!\!\times\!\!10^{8}$ & 54.7 & $7\!\!\times\!\!10^{6}$ - $7\!\!\times\!\!10^{9}$ \\ 
Sgr A* & 266.4 & -29.0 & 2.1 & 220.8 & $1\!\!\times\!\!10^{6}$ - $9\!\!\times\!\!10^{8}$ & 27.3 & $8\!\!\times\!\!10^{6}$ - $7\!\!\times\!\!10^{9}$ \\
 PKS 1622-253 & 246.4 & -25.5 & 2.1 & 184.2 & $1\!\!\times\!\!10^{6}$ - $8\!\!\times\!\!10^{8}$ & 23.3 & $7\!\!\times\!\!10^{6}$ - $6\!\!\times\!\!10^{9}$ \\
 1830-210 & 278.4 & -21.1 & 2.1 & 120.3 & $8\!\!\times\!\!10^{5}$ - $6\!\!\times\!\!10^{8}$ & 17.8 & $4\!\!\times\!\!10^{6}$ - $7\!\!\times\!\!10^{9}$ \\
 1730-130  & 263.3 & -13.1 & 4.7 & 91.5 & $2\!\!\times\!\!10^{5}$ - $3\!\!\times\!\!10^{8}$ & 20.5 & $1\!\!\times\!\!10^{6}$ - $6\!\!\times\!\!10^{9}$ \\
 PKS 1510-089 & 228.2 & -9.1 & 3.5 & 38.1 & $8\!\!\times\!\!10^{4}$ - $2\!\!\times\!\!10^{8}$ & 10.4 & $8\!\!\times\!\!10^{5}$ - $6\!\!\times\!\!10^{9}$ \\
 PKS 1406-076 & 212.2 & -7.9 & 4.1 & 36.5 & $6\!\!\times\!\!10^{4}$ - $2\!\!\times\!\!10^{8}$ & 10.6 & $6\!\!\times\!\!10^{5}$ - $6\!\!\times\!\!10^{9}$ \\
 2022-077 & 306.4 & -7.6 & 2.7 & 23.1 & $6\!\!\times\!\!10^{4}$ - $2\!\!\times\!\!10^{8}$ &  6.8 & $6\!\!\times\!\!10^{5}$ - $6\!\!\times\!\!10^{9}$ \\
 3C279  & 194.1 & -5.8 & 3.9 & 25.5 & $5\!\!\times\!\!10^{4}$ - $1\!\!\times\!\!10^{8}$ &  8.2 & $5\!\!\times\!\!10^{5}$ - $6\!\!\times\!\!10^{9}$ \\
 0336-019  & 54.9 & -1.8 & 2.9 & 13.7 & $3\!\!\times\!\!10^{4}$ - $9\!\!\times\!\!10^{7}$ &  5.5 & $3\!\!\times\!\!10^{5}$ - $5\!\!\times\!\!10^{9}$ \\
 PKS 0420-014 & 65.8 & -1.3 & 3.1 & 14.3 & $2\!\!\times\!\!10^{4}$ - $8\!\!\times\!\!10^{7}$ &  5.9 & $3\!\!\times\!\!10^{5}$ - $5\!\!\times\!\!10^{9}$ \\
3C 273  & 187.3 &  2.0 & 1.5 &  - \footnotemark[1] & $2\!\!\times\!\!10^{4}$ - $3\!\!\times\!\!10^{7}$ &  - \footnotemark[1] & $2\!\!\times\!\!10^{5}$ - $1\!\!\times\!\!10^{9}$ \\
 4C+10.45  & 242.2 & 10.5 & 3.3 & 11.5 & $2\!\!\times\!\!10^{4}$ - $5\!\!\times\!\!10^{6}$ & 14.1 & $6\!\!\times\!\!10^{4}$ - $5\!\!\times\!\!10^{7}$ \\
 PKS 1502+106 & 226.1 & 10.5 & 1.5 &  5.2 & $2\!\!\times\!\!10^{4}$ - $5\!\!\times\!\!10^{6}$ &  6.4 & $6\!\!\times\!\!10^{4}$ - $5\!\!\times\!\!10^{7}$ \\
 CTA 102  & 338.1 & 11.7 & 3.3 & 12.0 & $2\!\!\times\!\!10^{4}$ - $4\!\!\times\!\!10^{6}$ & 15.8 & $6\!\!\times\!\!10^{4}$ - $4\!\!\times\!\!10^{7}$ \\
M87 & 187.7 & 12.4 & 1.7 &  - \footnotemark[1] & $2\!\!\times\!\!10^{4}$ - $4\!\!\times\!\!10^{6}$ &  - \footnotemark[1] & $5\!\!\times\!\!10^{4}$ - $4\!\!\times\!\!10^{7}$ \\
PKS 0528+134 & 82.7 & 13.5 & 1.5 &  - \footnotemark[1] & $2\!\!\times\!\!10^{4}$ - $4\!\!\times\!\!10^{6}$ &  - \footnotemark[1] & $5\!\!\times\!\!10^{4}$ - $3\!\!\times\!\!10^{7}$ \\
3C 454.3  & 343.5 & 16.1 & 3.5 & - \footnotemark[1] & $1\!\!\times\!\!10^{4}$ - $4\!\!\times\!\!10^{6}$ & - \footnotemark[1] & $4\!\!\times\!\!10^{4}$ - $2\!\!\times\!\!10^{7}$ \\
 PKS 0235+164  & 39.7 & 16.6 & 3.5 & 16.0 & $1\!\!\times\!\!10^{4}$ - $4\!\!\times\!\!10^{6}$ & 24.6 & $4\!\!\times\!\!10^{4}$ - $2\!\!\times\!\!10^{7}$ \\
 OJ 248  & 127.7 & 24.2 & 3.5 & 18.7 & $1\!\!\times\!\!10^{4}$ - $2\!\!\times\!\!10^{6}$ & 37.7 & $3\!\!\times\!\!10^{4}$ - $1\!\!\times\!\!10^{7}$ \\
 0430+2859 & 68.4 & 29.1 & 3.7 & 23.6 & $9\!\!\times\!\!10^{3}$ - $2\!\!\times\!\!10^{6}$ & 52.3 & $2\!\!\times\!\!10^{4}$ - $7\!\!\times\!\!10^{6}$ \\
 OS 319  & 243.4 & 34.2 & 4.9 & 32.4 & $7\!\!\times\!\!10^{3}$ - $1\!\!\times\!\!10^{6}$ & 79.6 & $2\!\!\times\!\!10^{4}$ - $7\!\!\times\!\!10^{6}$ \\
4C +38.41  & 248.8 & 38.1 & 3.3 & - \footnotemark[1] & $7\!\!\times\!\!10^{3}$ - $1\!\!\times\!\!10^{6}$ & - \footnotemark[1] & $2\!\!\times\!\!10^{4}$ - $6\!\!\times\!\!10^{6}$ \\
 4C+51.37  & 265.1 & 52.2 & 2.9 & 23.8 & $4\!\!\times\!\!10^{3}$ - $7\!\!\times\!\!10^{5}$ & 83.2 & $1\!\!\times\!\!10^{4}$ - $2\!\!\times\!\!10^{8}$ \\
S5 0716+714 & 110.5 & 71.3 & 1.3 & - \footnotemark[1] & $2\!\!\times\!\!10^{3}$ - $3\!\!\times\!\!10^{5}$ & - \footnotemark[1] & $7\!\!\times\!\!10^{3}$ - $9\!\!\times\!\!10^{7}$ \\
\botrule

 \end{tabular} \label{sourcelist}}
 \end{table}

\end{document}